\documentclass[twocolumn,english,showpacs,preprintnumbers,superscriptaddress,nofootinbib,prbr]{revtex4-1}
\usepackage[latin9]{inputenc}
\usepackage[english]{babel}
\usepackage{setspace}
\usepackage{graphicx}
\usepackage{amsmath}
\usepackage{amssymb}
\usepackage{amscd}
\usepackage{rotating}
\usepackage[usenames, dvipsnames]{color}
\usepackage{epstopdf}
\usepackage{comment}
\usepackage{siunitx}
\usepackage{float}
\usepackage{xfrac}
\usepackage{natbib}
\usepackage{hyperref}
\hypersetup{
    colorlinks=true,
    linkcolor=blue,
    filecolor=blue,
    urlcolor=blue,
   citecolor=blue,
}
\usepackage[draft]{todonotes}

\usepackage[normalem]{ulem}
\newcommand\footnoteref[1]{\protected@xdef\@thefnmark{\ref{#1}}\@footnotemark}

\begin{document}

\title{Chiral magnetic chemical bonds in molecular states of impurities in Weyl semimetals}

\author{Y. Marques}
\email{yurimarques111@gmail.com}
\affiliation{Departamento de F\'{i}sica e Qu\'{i}mica, Universidade Estadual Paulista (Unesp), Faculdade de Engenharia, 15385-000, Ilha Solteira, S\~ao Paulo, Brazil}
\author{W. N. Mizobata}
\affiliation{Departamento de F\'{i}sica e Qu\'{i}mica, Universidade Estadual Paulista (Unesp), Faculdade de Engenharia, 15385-000, Ilha Solteira, S\~ao Paulo, Brazil}
\author{R. S. Oliveira}
\affiliation{Departamento de F\'{i}sica e Qu\'{i}mica, Universidade Estadual Paulista (Unesp), Faculdade de Engenharia, 15385-000, Ilha Solteira, S\~ao Paulo, Brazil}
\author{M. de Souza}
\affiliation{Departamento de F\'{i}sica, Instituto de Geoci\^encias e Ci\^encias Exatas, Universidade Estadual Paulista (Unesp), 13506-970, Rio Claro, S\~ao Paulo, Brazil}
\author{M. S. Figueira}
\affiliation{Instituto de F\'{i}sica, Universidade Federal Fluminense, 24210-340, Niterói, RJ, Brazil}
\author{I. A. Shelykh}
\affiliation{Science Institute, University of Iceland, Dunhagi-3, IS-107,
Reykjavik, Iceland}
\affiliation{ITMO University, St. Petersburg 197101, Russia}
\author{A. C. Seridonio}
\email[correspondent author: ]{antonio.seridonio@unesp.br}
\affiliation{Departamento de F\'{i}sica e Qu\'{i}mica, Universidade Estadual Paulista (Unesp), Faculdade de Engenharia, 15385-000, Ilha Solteira, S\~ao Paulo, Brazil}
\affiliation{Departamento de F\'{i}sica, Instituto de Geoci\^encias e Ci\^encias Exatas, Universidade Estadual Paulista (Unesp), 13506-970, Rio Claro, S\~ao Paulo, Brazil}

\date{\today}

\begin{abstract}
We demonstrate that chirality of the electron scattering in Weyl semimetals leads to the formation of magnetic chemical bonds for molecular states of a pair of impurities. The effect is associated with the presence of time-reversal symmetry breaking terms in the Hamiltonian which drive a crossover from \textit{s-} to \textit{p-}wave scattering. The profiles of the corresponding molecular orbitals and their spin polarizations are defined by the relative orientation of the lines connecting two Weyl nodes and two impurities. The magnetic character of the molecular orbitals and their tunability open the way for using doped Weyl semimetals for \textit{spintronics} and realization of \textit{qubits}.
\end{abstract}
\maketitle

\textit{Introduction.}~Recent years witnessed unprecedented penetration of the ideas of high energy physics into the domain of condensed matter. In particular, lot of attention is now attracted to the condensed matter realizations of three dimensional (3D) massless  quasi-relativistic particles known as Dirac or Weyl fermions\cite{Weyl1929}. The experimental observation of Dirac fermions in such materials as Na$_3$Bi\cite{Wang2012,Liu2014a} and Cd$_3$As$_2$\cite{Wang2013,Liu2014b} made possible the study of the 3D analogs of graphene physics in a robust topologically protected material possessing both inversion ($\mathcal{I}$) and time reversal ($\mathcal{T}$) symmetries \cite{Armitage2018}. In Weyl semimetals, where one of these symmetries is broken, a Dirac node, which is the point where conduction and valence bands touch each other, splits into a pair of Weyl nodes with opposite chiralities. Such nodes are predicted to give rise to a plethora of interesting phenomena, including formation of Fermi arcs, unusual Hall effects, and chiral anomaly, among others\cite{Armitage2018,Wan2011,Yang2011,Hosur2012,Kim2017,Nielsen1983,GXu2011}. The material platform for realization of Weyl fermions is presented by such compounds as tantalum arsenide (TaAs)\cite{Huang2015,Weng2015,Xu2015I,Lv2015I,Lv2015II}, niobium arsenide (NbAs)\cite{Xu2015II}, and tantalum phosphide (TaP)\cite{Xu2016}.

One of the aspects of Weyl semimetals which recently received particular attention is the peculiar impurity physics\cite{Sun2015,Ma2015,Chang2015,Principi2015,Zheng2016,Marques2017}. In the present work, we clarify the role played by chirality of Weyl quasiparticles in the processes of impurity scattering by investigation of the local density of states. The latter can be experimentally addressed by means of the scanning electron microscopy (STM). We show that long-range \textit{Friedel-like oscillations}\cite{Friedel} contribute to the formation of molecular states in a pair of distant impurities embedded in a 3D relativistic semimetal. We demonstrate that, the scenario of the impurity scattering is radically different in Dirac and Weyl semimetals and show that in the latter case  magnetic molecular states can be formed. Their particular type is defined by the relative orientation of the lines connecting two Weyl nodes and two impurities. We report a crossover from \textit{s-} to \textit{p-}type atomic orbitals for individual impurities and related formation of spin-polarized $\sigma-$ and $\pi-$type molecular orbitals for an impurity pair.

\begin{figure}[!]
	\centering\includegraphics[width=2.0in,keepaspectratio]{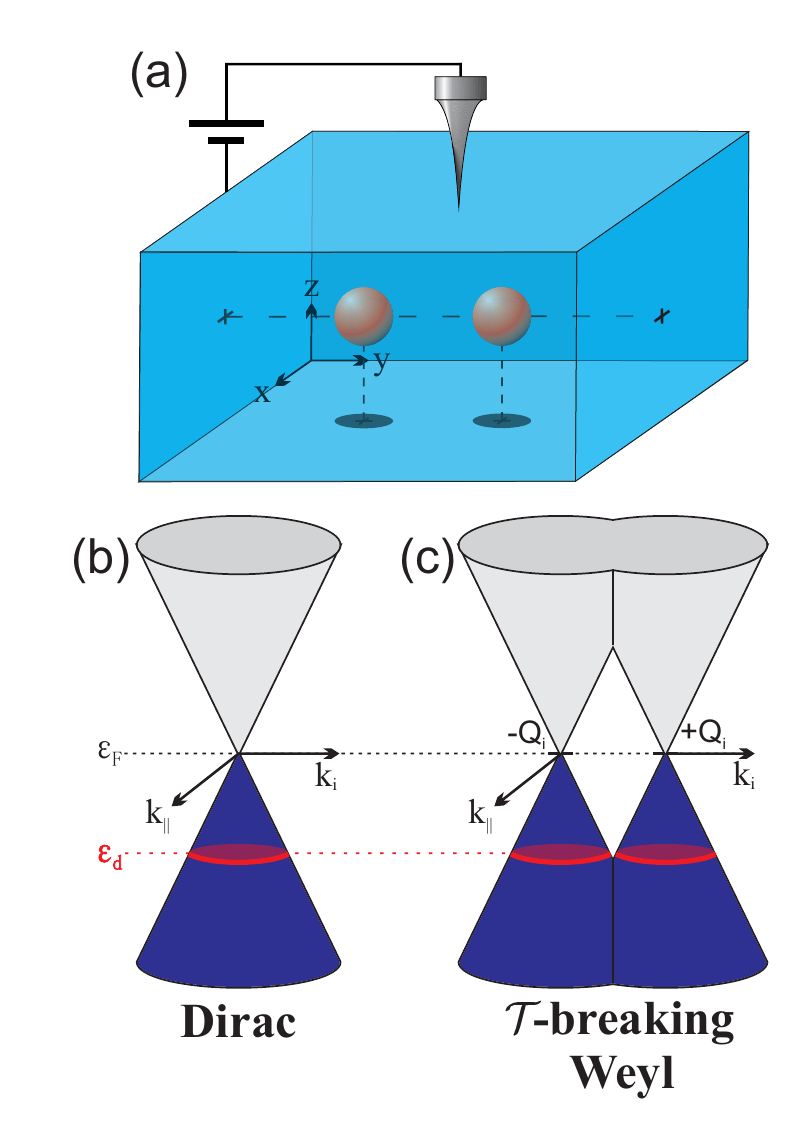}\protect\protect\protect\caption{\label{fig:Pic1} {(Color online) Panel (a): Sketch of the proposed setup. Two impurities are embedded in a 3D semimetal of Dirac or Weyl type. The density of electrons forming molecular orbitals can be probed by an STM-tip. Panels (b), (c) show low energy band structure for Dirac and $\mathcal{T}$-breaking Weyl semimetals with two Weyl nodes located at $\pm Q_{i}$, $i=x,y,z$. The blue color of the lower cones indicates the filling of the valence bands, black dotted line is the Fermi energy set at $\varepsilon_{F}=0$ and red dotted line corresponds to the single-particle energy of the impurities.}}
\end{figure}
\textit{The Model.} We set $\hbar=1$ throughout the calculations and represent the total Hamiltonian as the sum of the three terms:
\begin{equation}
\mathcal{H}=\mathcal{H}_{\text{{0}}}+\mathcal{H}_{\text{{d}}}+\mathcal{H}_{\mathcal{V}.}
\end{equation}

The low-energy Hamiltonian of the host may be represented as
\begin{eqnarray}
\mathcal{H}_{\text{{0}}} & = & \sum_{\mathbf{k}}\psi^{\dagger}(\mathbf{k})(H_{+}\oplus H_{-})\psi(\mathbf{k}),\label{eq:band-1}
\end{eqnarray}
where $\psi(\mathbf{k})=(c_{\mathbf{k}+\uparrow},c_{\mathbf{k}+\downarrow},c_{\mathbf{k}-\uparrow},c_{\mathbf{k}-\downarrow})^{T}$
is four-spinor operator whose components $c_{\mathbf{k}\chi\sigma}^{\dagger}$
($c_{\mathbf{k}\chi\sigma}$) stand for the creation (annihilation)
operators of an electron with wave number $\boldsymbol{k}$ and spin $\sigma$,
\begin{eqnarray}
H_{\chi}(\mathbf{k}) & = & \chi v_{F}\boldsymbol{\sigma}\cdot(\mathbf{k}-\chi\boldsymbol{Q}),\label{eq:ht-1}
\end{eqnarray}
where $\mathbf{k}=(k_{x},k_{y},k_{z})$ is the three-dimensional wave vector, $\boldsymbol{\sigma}$ stands for the vector of Pauli matrices, the index $\chi=\pm1$ corresponds to the chirality of the Weyl nodes and $v_{F}$ is the Fermi velocity. For $\boldsymbol{Q}=\boldsymbol{0}$, $\mathcal{T}$ symmetry is conserved and a pair of Weyl nodes is degenerated, which corresponds to the case of a standard Dirac semimetal. If $\mathcal{T}$ symmetry is broken ($\boldsymbol{Q}\neq\boldsymbol{0}$), two Weyl nodes are displaced with respect to each other towards two different points in the Brillouin zone located at $\pm\boldsymbol{Q}$, but maintain energetic degeneracy as it is depicted in the Fig. (\ref{fig:Pic1}).

The impurities are modeled by the Hamiltonian
\begin{eqnarray}
\mathcal{H}_{\text{{d}}} & = & \sum_{j\sigma}\varepsilon_{j\sigma}d_{j\sigma}^{\dagger}d_{j\sigma}+\sum_{j}U_{j}n_{j\uparrow}n_{j\downarrow},\label{eq:Imp-1}
\end{eqnarray}
with $\varepsilon_{j\sigma}$ being single-particle energy and $U_{j}$ the on-site Coulomb repulsion, whereas  $n_{j\sigma}=d_{j\sigma}^{\dagger}d_{j\sigma}$
corresponds to the number of electrons with spin projection $\sigma$ at the site
$j$ with $d_{j\sigma}^{\dagger}$ and $d_{j\sigma}$ being respectively creation and annihilation operators.

The hybridization between the host and the impurities is described by the term:
\begin{eqnarray}
\mathcal{H}_{\mathcal{V}} & = & \sum_{j\mathbf{k}}\hat{d}_{j}^{\dagger}\hat{V}_{j \mathbf{k}}\psi(\mathbf{k})+\text{H.c.},\label{eq:Hibr-1}
\end{eqnarray}
wherein $\hat{d}_{j}^{\dagger}=(\begin{array}{cc}
d_{j\uparrow}^{\dagger}, & d_{j\downarrow}^{\dagger})\end{array}$ and
\begin{eqnarray}
\hat{V}_{j \mathbf{k}} & = & \left(\begin{array}{cc}
V_{j\mathbf{k}} & 0\\
0 & V_{j\mathbf{k}}
\end{array}\begin{array}{cc}
V_{j\mathbf{k}} & 0\\	0 & V_{j\mathbf{k}}
\end{array}\right),\label{eq:Vkmatrix-1}
\end{eqnarray}
where
$V_{j\mathbf{k}}=\frac{v_{0}}{\sqrt{N}}e^{i\mathbf{k}\cdot\mathbf{R}_{j}}$, with $v_{0}$ being the hybridization amplitude between electrons of the host and localized states of the impurities positioned at $\mathbf{R}_{j}$ ($j=1,2$), $N$ is the normalization factor yielding the total number of the conduction states.
\textit{Local Density of States (LDOS).} The electronic properties of the considered system are determined by the LDOS of the host which can be experimentally accessed by means of an STM-tip. It can be calculated using standard equation-of-motion (EOM) procedure\cite{Haug1996,Bruus} as:
\begin{equation}
\rho(\varepsilon,{\bold r_{m}})=-\frac{1}{\pi}\sum_\sigma\textrm{Im}\{\tilde{\mathcal{G}}_\sigma(\varepsilon,{\bold r_{m}})\}=\rho_{0}(\varepsilon)+\sum_{jj'}\delta\rho_{jj'}(\varepsilon,{\bold r_{m}}),\label{eq:ldos}
\end{equation}
where $\tilde{\mathcal{G}}_\sigma(\varepsilon,{\bold r_{m}})$ is the time-Fourier transform of the retarded Green\textquoteright s function in the time domain, defined as:
\begin{eqnarray}
\mathcal{G}_\sigma(t,{\bold r_{m}}) & = & -i\theta\left(t\right)\left\langle \{\psi_{\sigma}(t,{\bold r_{m}}),\psi_{\sigma}^{\dagger}(0,{\bold r_{m}})\}\right\rangle _{\mathcal{H}},
\end{eqnarray}
where $\theta\left(t\right)$ denotes the Heaviside function, $\psi_{\sigma}(t,{\bold r_{m}})$ is the field operator of the host electrons written in terms of the continuous variable $\mathbf{r}_{m}$, the brackets $\left\langle \cdots\right\rangle _{\mathcal{H}}$ denote the ensemble average with respect to the full Hamiltonian, $\{\cdots\}$ determines an anti-commutator between operators in the Heisenberg picture, $\rho_{0}(\varepsilon)=\frac{6\varepsilon^{2}}{D^{3}}$ is the pristine host DOS with $D$ being the energy cutoff corresponding to the half-bandwidth, and
\begin{equation}
\delta\rho_{jj'}(\varepsilon)=-\frac{1}{\pi v_{0}^{2}}\sum_{\chi\chi'}\sum_{\sigma}\textrm{Im}\Bigl[\Sigma_{\sigma}^{\chi}\bigl({\bold r_{mj}}\bigr){\cal {\cal \tilde{G}}}_{j\sigma|j'\sigma}(\varepsilon)\Sigma_{\sigma}^{\chi'}\bigl({\bold r_{j'm}}\bigr)\Bigr]\label{eq:dldos}
\end{equation}
encodes the \textit{Friedel-like oscillations} describing the scattering of the conduction electrons by the impurities, where the terms $j'=j$ and $j'\neq{j}$ give rise to intra and inter-impurity scattering processes, respectively, which are ruled by the spatial dependent self-energy
\begin{eqnarray}
\Sigma_{\sigma}^{\chi}({\bold r_{mj}}) & = & -\frac{\xi\pi v_{F}}{D|{\bold r_{mj}}|}e^{-i|{\bold r_{mj}}|\frac{\varepsilon}{v_{F}}}e^{\mp i\chi\boldsymbol{Q}\cdot{\bold r_{mj}}}\nonumber \\
& \times & \Bigl(\varepsilon\pm\chi\sigma\varepsilon\pm i\frac{\sigma v_{F}\chi}{|{\bold r_{mj}}|}\Bigr),\label{eq:SEr}
\end{eqnarray}
responsible for the chiral magnetic chemical bound mechanism as we will see below, ${\bold r_{mj}}={\bold r_{m}}-{\bold R_{j}}$ and $\pm$ signs correspond to the vector direction (positive for ${\bold r_{mj}}$, negative for ${\bold r_{jm}}$), $\tilde{G}_{j\sigma|j'\sigma}(\varepsilon)$ is the time-Fourier transform of the Green's function of the impurities
\begin{equation}
\mathcal{G}_{j\sigma|j'\sigma}=-i\theta\left(t\right)\bigl\langle\{d_{j\sigma}\left(t\right),d_{j'\sigma}^{\dagger}\left(0\right)\}\bigr\rangle_{\mathcal{H}}.
\end{equation}

Application of the EOM method to $\tilde{G}_{j\sigma|j'\sigma}(\varepsilon)$ together with Hubbard-I decoupling scheme\cite{HubbardI}, yields
\begin{eqnarray}
\mathcal{G}_{j\sigma|j\sigma}(\varepsilon) & = & \frac{\lambda_{j}^{\bar{\sigma}}}{g_{j\sigma|j\sigma}^{-1}(\varepsilon)-\lambda_{j}^{\bar{\sigma}}{{\Sigma}_{\sigma}({\bold r_{j{j'}}})g_{j'\sigma|j'\sigma}(\varepsilon)\lambda_{{j'}}^{\bar{\sigma}}{\Sigma}_{\sigma}({\bold r_{{j'}j}})}},\nonumber \\
\label{eq:Gjj}
\end{eqnarray}
where $\bar{\sigma}=-\sigma$, $j\neq{j'}$, ${\bold r_{jj'}}={\bold R_{j}}-{\bold R_{j'}}$, $\lambda_{j}^{\bar{\sigma}}=1+\frac{U_{j}}{g_{j\sigma|j\sigma}^{-1}(\varepsilon)-U_{j}}\bigl\langle n_{j\bar{\sigma}}\bigr\rangle$ is the spectral weight, $g_{j\sigma|j\sigma}(\varepsilon)=\frac{1}{\varepsilon-\varepsilon_{j\sigma}-{\Sigma}_{0}}$ as the single impurity noninteracting Green's function,
\begin{eqnarray}
\bigl\langle n_{j\bar{\sigma}}\bigr\rangle & = & -\frac{1}{\pi}\int_{-\infty}^{+\infty}n_{F}(\varepsilon)\textrm{Im}(\mathcal{G}_{j\bar{\sigma}|j\bar{\sigma}})d\varepsilon\label{eq:ON}
\end{eqnarray}
is the occupation number of an impurity with $n_{F}(\varepsilon)$ being the Fermi-Dirac distribution,
\begin{eqnarray}
{\Sigma}_{0} & = & \frac{3v_{0}^{2}}{D^{2}}\Bigl(\frac{\varepsilon^{2}}{D}\text{ln}\left|\frac{D+\varepsilon}{D-\varepsilon}\right|-2\varepsilon-i\frac{\varepsilon^{2}}{D}\Bigr)\label{eq:SE0}
\end{eqnarray}
is the local self-energy, ${\Sigma}_{\sigma}({\bold r_{j{j'}}})=\sum_{\chi}\Sigma_{\sigma}^{\chi}({\bold r_{j{j'}}})$ and
\begin{eqnarray}
\tilde{{\cal G}}_{d_{j\sigma}d_{{j'}\sigma}}\left(\varepsilon\right) & = & g_{j\sigma|j\sigma}(\varepsilon){\lambda_{j}^{\bar{\sigma}}{\Sigma}_{\sigma}({\bold r_{j{j'}}})}{\cal \tilde{{\cal G}}}_{d_{{j'}\sigma}d_{{j'}\sigma}}\left(\varepsilon\right).\label{eq:Gjl}
\end{eqnarray}
\textit{Results and Discussion.} In order to understand the formation of the molecular states of a pair of impurities inside a Weyl semimetal, we should start from analyzing the case of a single impurity. As model parameters, we adopt without loss of generality, temperature $T=0\text{K}$, the energy of an impurity $\varepsilon_{j\sigma}=-0.07D$, hybridization amplitude $v_{0}=-0.14D$, on-site Coulomb repulsion $U_{j}=0.14D$, $\hbar v_{F}\approx3\:eV\text{\AA}$ and $D\approx0.2\,\text{\text{{eV}}}$.

\begin{figure}
	\centering{}\includegraphics[width=3.5in,keepaspectratio]{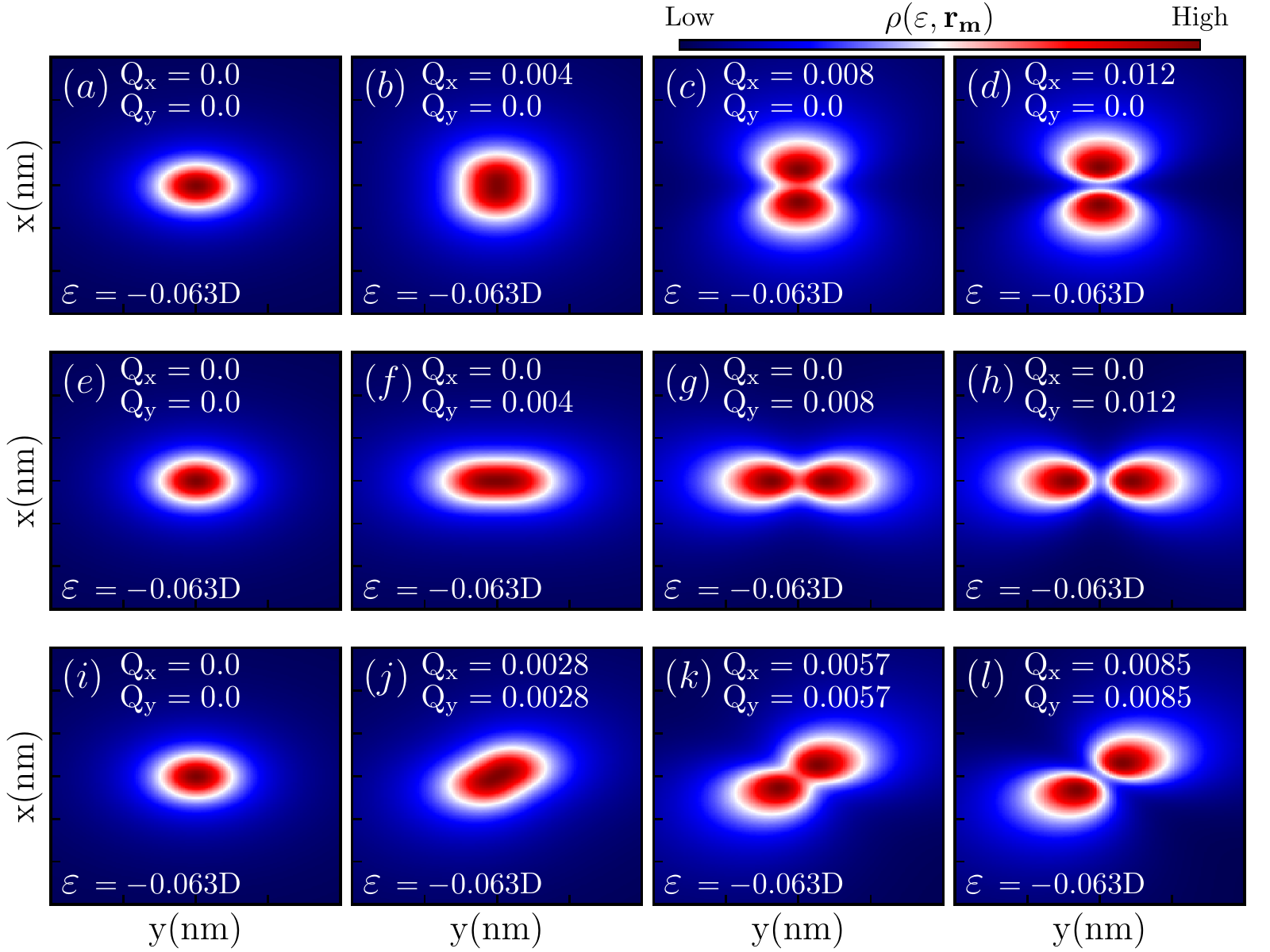}
	\protect\protect\protect\caption{\label{fig:SIAM}(Color online) 2D LDOS maps for the case of a single impurity taken at fixed energy $\varepsilon$. Crossover from $s-$ to $p-$type orbitals associated with moving from Dirac ($\boldsymbol{Q}=\boldsymbol{0}$) to Weyl ($\boldsymbol{Q}\boldsymbol{\neq0}$) regime is clearly seen.}
\end{figure}

As one can see in the Fig.\ref{fig:SIAM}, the 2D map of the LDOS which can be probed by an STM-tip over the system surface presents a crossover from \textit{s-} to \textit{p-}type atomic orbitals as $\boldsymbol{Q}$ is increased and one moves from Dirac ($\boldsymbol{Q}=\boldsymbol{0}$) towards the Weyl regime ($\boldsymbol{Q}\boldsymbol{\neq0}$). This happens due to the presence of the terms depending on $v_{F}\chi\boldsymbol{Q}$ in the original Hamiltonian. Note, that the \textit{p}-orbital is elongated along the direction of $\boldsymbol{Q}$.

\begin{figure}
	\centering{}\includegraphics[width=3.5in,keepaspectratio]{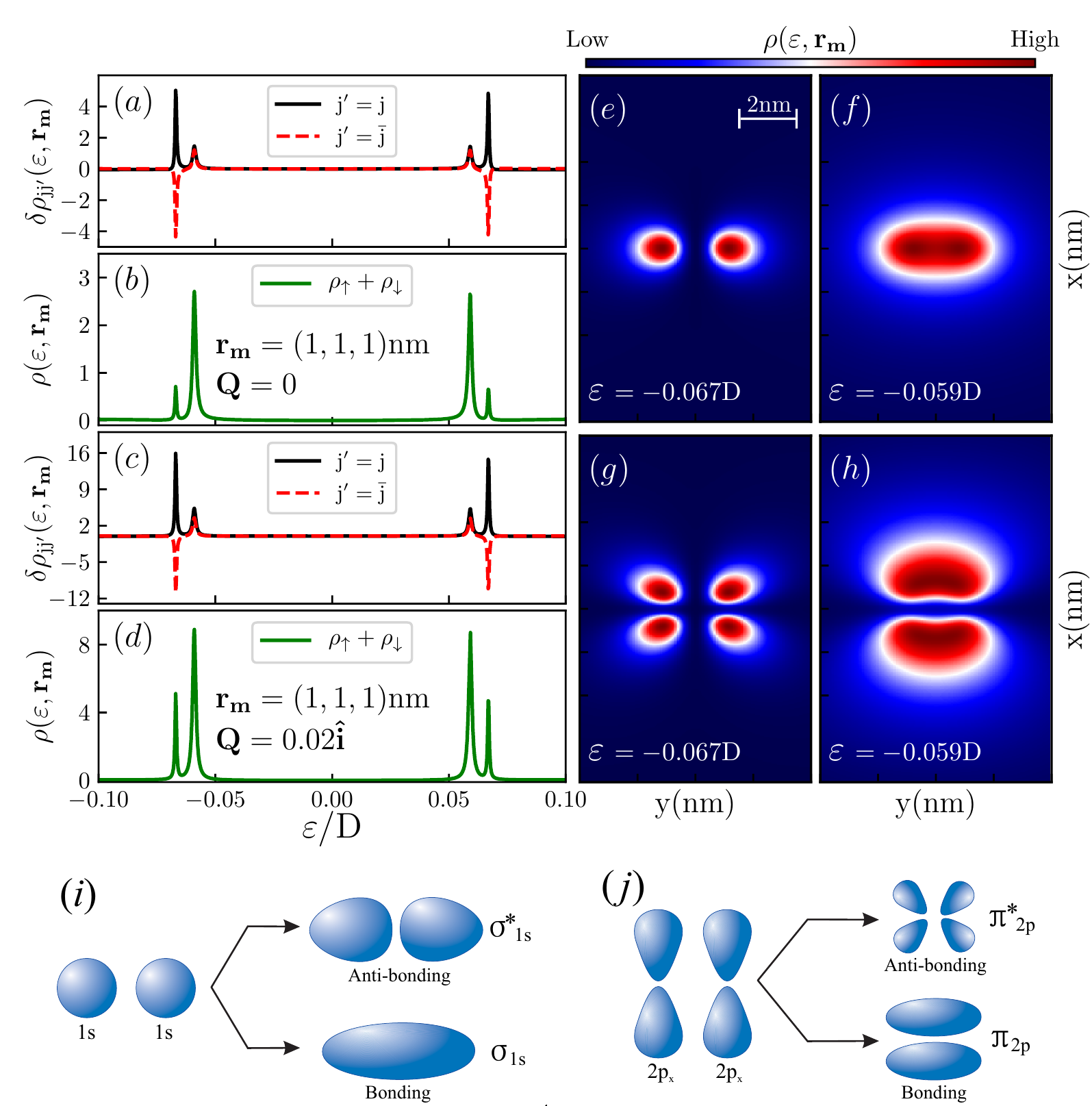}
	\protect\protect\caption{\label{fig:Pic2} (Color online) LDOS for a pair of impurities. The panels (a),(b),(e),(f)) correspond to the case of Dirac  semimetal ($\boldsymbol{Q}=0$), the panels (c),(d),(g),(h)) to the case of $\mathcal{T}$-breaking Weyl semimetal with $\boldsymbol{Q}\cdot\boldsymbol{r}_{12}=0$.
		Panels (a) and (c) display the diagonal ($\delta\rho_{jj}$) and off-diagonal ($\delta\rho_{j\bar{j}}$) contributions to the LDOS. Note that $\delta\rho_{j\bar{j}}$ reveals both dips and peaks corresponding to anti-resonances and resonances respectively, while $\delta\rho_{jj}$ reveals peaks only. The total LDOS is presented in the panels (b) and (d). We set $\mathbf{r}_{m}=(1,1,1)\text{{nm}}$, the energy is counted from the Fermi level set at $\varepsilon_{F}=0$, for the Weyl host $Q_{x}=0.02$. The total LDOS on the $\mathbf{r}_{m}=(x,y,1)\text{{nm}}$ surface for the energies of bonding and antibonding states ($\varepsilon=-0.067D$ and $\varepsilon=-0.059D$) is shown in panels (e,f,g,h). Panels (i) and (j) illustrate how molecular orbitals presented in the panels (e)-(h) are formed from atomic orbitals presented in the Fig.2.}
\end{figure}

Now we can analyze the molecular state corresponding to a pair of impurities inside a Weyl semimetal with broken $\mathcal{T}-$symmetry. We will consider two cases of the mutual orientation of the vectors $\boldsymbol{Q}$ and $\boldsymbol{r}_{12}=\boldsymbol{R}_{1}-\boldsymbol{R}_{2}$ connecting the two impurities: \textit{i)} perpendicular orientation, $\boldsymbol{Q}\cdot\boldsymbol{r}_{12}=0$ and \textit{ii)} parallel orientation $\boldsymbol{Q}\cdot\boldsymbol{r}_{12}=|\boldsymbol{Q}||\boldsymbol{r}_{12}|$. As we will demonstrate, the former case corresponds to the formation of spin degenerate molecular orbitals, while the latter case gives rise to the chiral magnetic chemical bonds.

Let us start from the case $\boldsymbol{Q}\cdot\boldsymbol{r}_{12}=0$. We place impurities in the XY plane inside the host at the positions corresponding to $\mathbf{R}_{1,2}=(0,\mp1,0)\text{{nm}}$. In the case of an individual impurity, a single energy resonance appears within the valence band in $\rho(\varepsilon,{\bold r_{m}})$. Naturally, in the two-impurity system a pair of peaks corresponding to bonding and antibonding states appears, as it is shown in the Figs.\ref{fig:Pic2}(b) and \ref{fig:Pic2}(d) for the cases of Dirac ($\boldsymbol{Q}=\boldsymbol{0}$) and $\mathcal{T}$-breaking Weyl ($\boldsymbol{Q}\neq\boldsymbol{0}$) hosts. Note, that the coupling between the impurities is fully mediated by \textit{Friedel-like oscillations} of the electronic density of the host mathematically described by the self-energy $\lambda_{j}^{\bar{\sigma}}{{\Sigma}_{\sigma}({\bold r_{j{\bar{j}}}})g_{\bar{j}\sigma|\bar{j}\sigma}(\varepsilon)\lambda_{{\bar{j}}}^{\bar{\sigma}}{\Sigma}_{\sigma}({\bold r_{{\bar{j}}j}})}$ entering into the denominator of $\tilde{{\cal G}}_{d_{j\sigma}d_{{{j}}\sigma}}\left(\varepsilon\right)$ given by the Eq.(\ref{eq:Gjj}).

\begin{figure}
	\centering{}\includegraphics[width=3.5in,keepaspectratio]{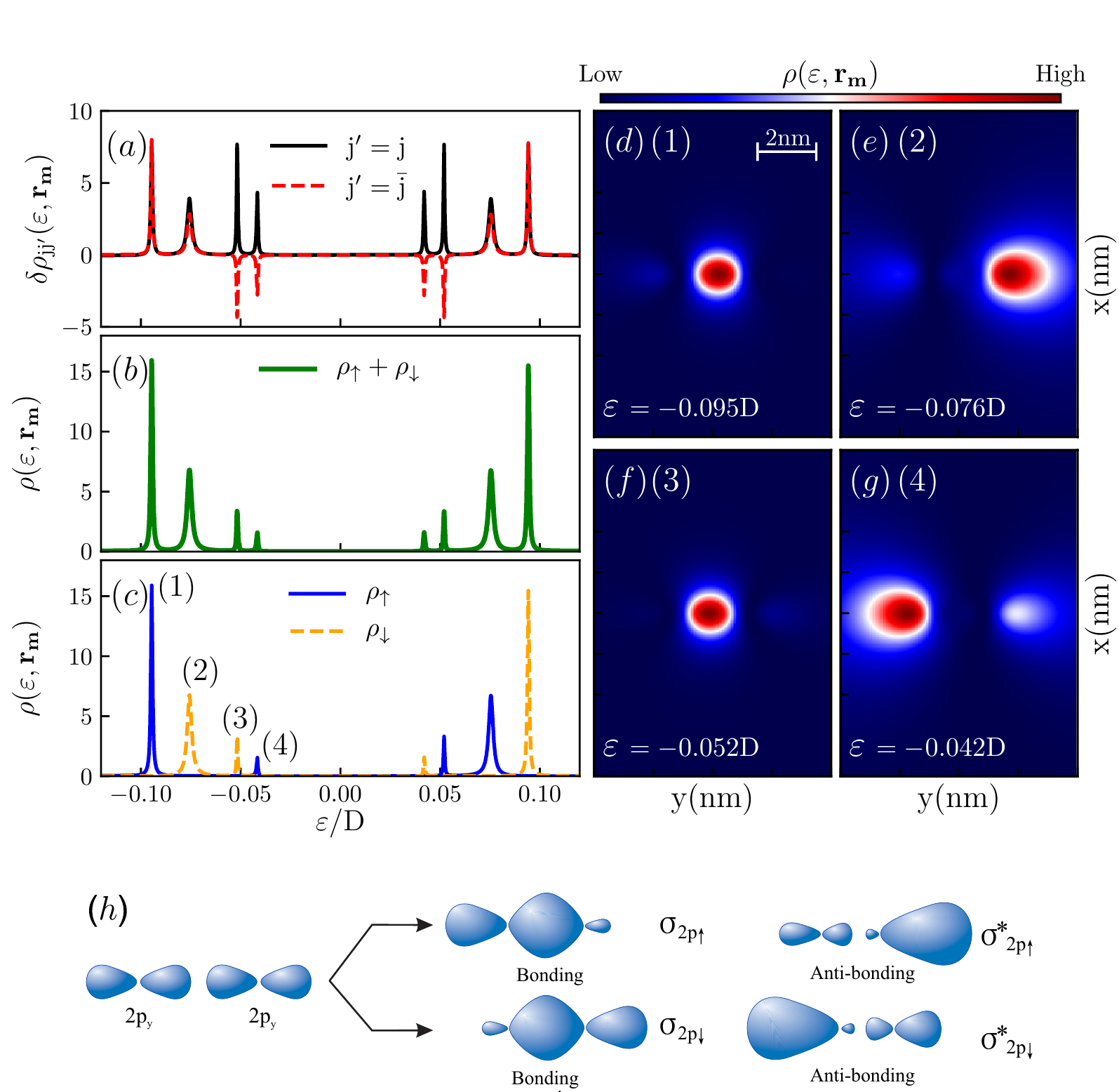}
	\protect\protect\protect\caption{\label{fig:Pic3}(Color online) LDOS for the $\mathcal{T}$-breaking Weyl semimetal for the case when vectors $\boldsymbol{Q}$ and $\boldsymbol{r}_{12}$ are parallel (we took $\boldsymbol{Q}=0.02\boldsymbol{\hat{j}}$)). Panel (a) displays diagonal ($\delta\rho_{jj}$) and off-diagonal ($\delta\rho_{j\bar{j}}$) contributions to the LDOS. The total LDOS is presented in the panel (b). Panel (c) shows spin-resolved density of states. The map of the total LDOS on the $\mathbf{r}_{m}=(x,y,1)\text{{nm}}$ surface for the energies corresponding to the four spin-resolved molecular states is presented in the panels (d)-(g). Panel (h) illustrates how molecular orbitals presented in the panels (d)-(g) are formed from atomic orbitals presented in the Fig.2.}
\end{figure}

The 2D map of the molecular orbitals on the host surface is presented in the panels (e)-(h) of the Fig.\ref{fig:Pic2}. Panels (e) and (f) correspond to the case of a Dirac host which was previously considered by some of us in the Ref.~\citenum{Marques2017}. One clearly sees the emergence of bonding and antibonding molecular orbitals with $\sigma$-type symmetry resulting from the interference between two  \textit{s-wave} atomic orbitals of individual impurities, as it is illustrated in the panel (i). Panels (g) and (h) correspond to the case of a Weyl semimetal with $\boldsymbol{Q}\cdot{\bold r_{j\bar{j}}}=0$, for which individual impurities reveal $p-$type atomic orbitals stretched in the direction perpendicular to the line connecting the impurities. Note, that for the considered case bonding and antibonding molecular orbitals have clear $\pi$-type symmetry. These orbitals remain spin-degenerate, as it follows from the Eq.(\ref{eq:SEr}), which leads to ${\Sigma}_{\sigma}({\bold r_{j{j'}}})=\sum_{\chi}\Sigma_{\sigma}^{\chi}({\bold r_{j{j'}}})$ independent of the spin degree of freedom, i.e., ${\Sigma}_{\uparrow}({\bold r_{j{j'}}})={\Sigma}_{\downarrow}({\bold r_{j{j'}}})$.

\begin{figure}
	\centering{}\includegraphics[width=3.5in,keepaspectratio]{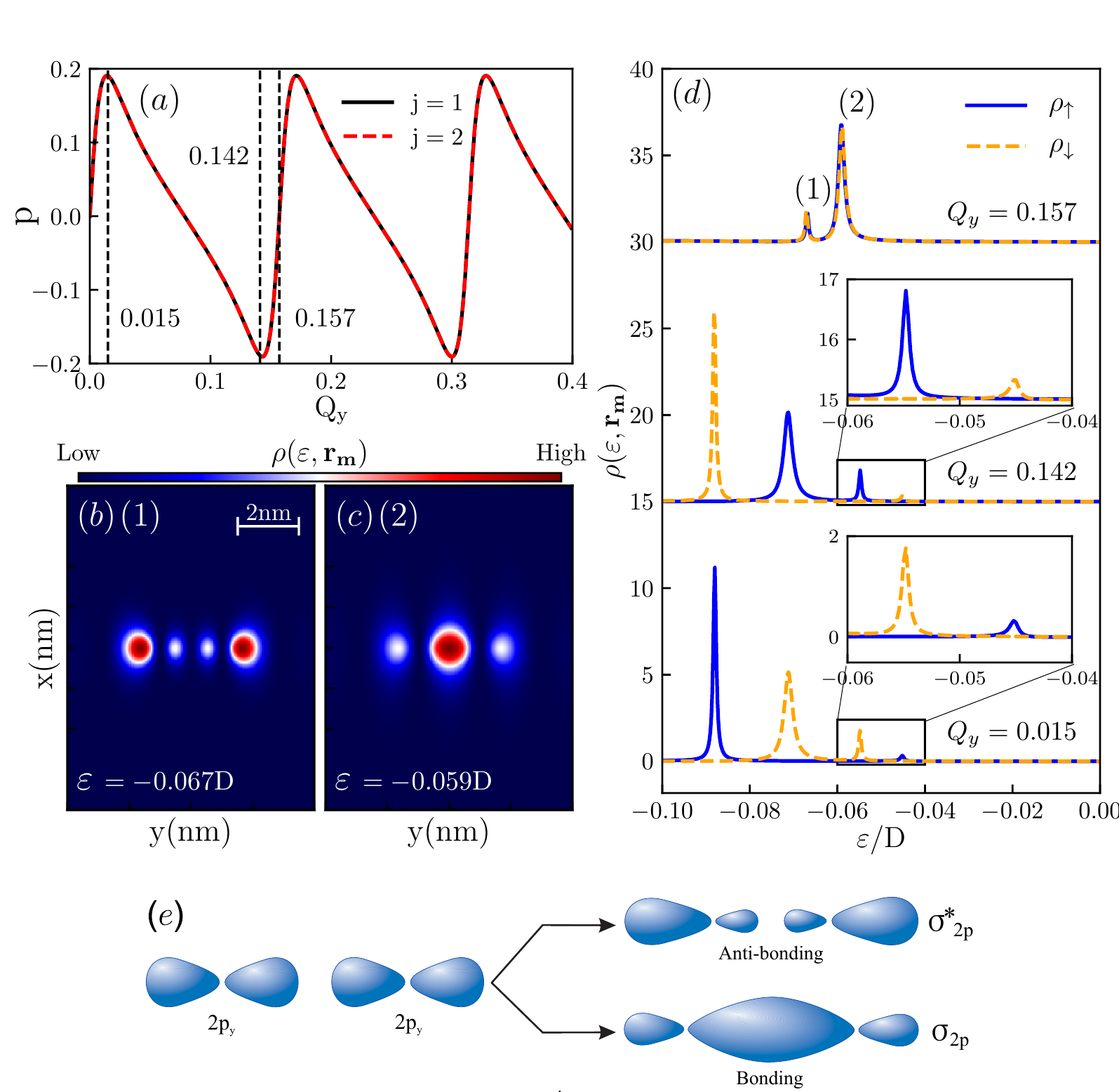}
	\protect\protect\protect\caption{\label{fig:Pic4}(Color online) Panel (a): Total magnetization of the impurities as function of the parameter $Q_{y}$ describing the shift of the Weyl nodes in the reciprocal space in the direction parallel to the line connecting impurities. Panels (b),(c): The maps of the total LDOS on the $\mathbf{r}_{m}=(x,y,1)\text{{nm}}$ surface for the energies corresponding to bonding and antibonding states in the valence band for the spin degenerate case corresponding to $Q_{y}=0.157$. Panel (d): Spin resolved LDOS at the point $\mathbf{r}_{m}=(1,1,1)\text{{nm}}$ for spin degenerate case ($Q_{y}=0.157$) and maximal spin up ($Q_{y}=0.015$) and spin down ($Q_{y}=0.142$) polarizations. The LDOS for $Q_{y}=0.157$ is rescaled by the factor of 5. {The insets highlight the marked sectors.} Panel (e)  illustrates how molecular orbitals presented in the panels (b) and (c) are formed from atomic orbitals presented in the Fig.2.}
\end{figure}

The case of the parallel orientation of the vectors $\boldsymbol{Q}$ and $\bold{ r}_{12}$  is illustrated by the Fig.\ref{fig:Pic3}. Note that in this case, according to the Eq. (\ref{eq:SEr}) the presence of the terms $e^{i\chi\boldsymbol{Q}\cdot{\bold r_{j\bar{j}}}}$ with $\chi=\pm1$ in the expression for the self-energy ${\Sigma}_{\sigma}({\bold r_{j{\bar{j}}}})=\sum_{\chi}\Sigma_{\sigma}^{\chi}({\bold r_{j{\bar{j}}}})$, leads to the lifting of spin degeneracy and gives rise to the formation of chiral magnetic chemical bonds. Interestingly enough, this spin-dependency can not be considered as being fully equivalent to one induced by effective external magnetic field, once the sequence of the peaks in the LDOS presented in the Fig.\ref{fig:Pic3}(b) does not correspond to the alternation of spin-up and spin-down states as usual, but consists of the two inner spin-down states flanked by the two outer spin-up states as can be clearly seen from the Fig.\ref{fig:Pic3}(c). The profiles of the spin-resolved orbitals corresponding to the bonding and antibonding states are shown in the panels (d)-(g) of the Fig.\ref{fig:Pic3}. These orbitals exhibit $\sigma$-type symmetry and are formed due to the interference between two frontal \textit{p-wave} orbitals as sketched in the Fig.\ref{fig:Pic3}(h).

To shed more light on the splitting between spin-polarized components in the LDOS, we investigate the impurity magnetization characterized by the polarization degree $p = (\bigl\langle n_{j\uparrow}\bigr\rangle-\bigl\langle n_{j\downarrow}\bigr\rangle)/\bigl\langle n_{j\uparrow}\bigr\rangle+\bigl\langle n_{j\downarrow}\bigr\rangle)$, where the occupation numbers are defined by the Eq.(\ref{eq:ON}). The dependence of the magnetization on the separation between the Weyl nodes in the direction parallel to the line connecting the two impurities $Q_{y}$ is shown in the Fig. \ref{fig:Pic4}(a). One clearly sees pronounced periodic behavior, stemming from the oscillations of the factor $e^{i\chi\boldsymbol{Q}\cdot{\bold r_{j\bar{j}}}}$ in the expression for spin-resolved self-energy in the Eq.(\ref{eq:SEr}). The LDOS corresponding to the spin-degenerate case,  maximal positive and negative magnetizations, is shown in the Fig. \ref{fig:Pic4}(d). Note that for the spin-degenerate situation corresponding to $Q_{y}=0.157$, the shape of the molecular orbitals presented in the  Figs.\ref{fig:Pic4}(b,c) can be represented as linear combination of the orbitals presented in the Fig.\ref{fig:Pic3}(d)-(g).

\textit{Conclusions.} We analyzed the structure of the molecular orbitals corresponding to the pair of impurities placed within a Weyl semimetal focusing on the role played by the $\mathcal{T}$-symmetry breaking. For this purpose the corresponding  LDOS was evaluated. It was demonstrated that the terms in the self-energy stemming from the chiral dependent minimal coupling drive a crossover from spin degenerate $\sigma$-type molecular orbitals characteristic for the case of a Dirac host to spin degenerate $\pi$-type orbitals or spin-polarized $\sigma$-type orbitals for the case of a Weyl host. The type of the chemical bonding in this latter case can be controlled by variation of the mutual position of the impurities with respect to the vector $\boldsymbol{Q}$ describing the shift of the Weyl nodes. The magnetic character of the molecular orbitals and their tunability open the way for using doped Weyl semimetals for \textit{spintronics} and realization of \textit{qubits}.

\textit{Acknowledgments.} We thank the funding Brazilian agencies CNPq (Grant No. 307573/2015-0), CAPES, and S\~ao Paulo Research Foundation (FAPESP) Grant No. 2018/09413-0. IAS acknowledges support from Horizon2020 project CoExAN,  megagrant 14.Y26.31.0015 and Goszadanie no. 3.8884.2017/8.9 of the Ministry of Education and Science of Russian Federation.

\end{document}